\documentclass[longauth]{aa}
\usepackage{graphicx}
\usepackage{txfonts}
\usepackage{xcolor}
\usepackage{hyperref}
\hypersetup{colorlinks, linkcolor=blue, citecolor=blue, urlcolor=blue}

\newcommand{\pivec}{\mbox{\boldmath $\pi$}}
\newcommand{\muvec}{\mbox{\boldmath $\mu$}}

\newcommand{\te}{t_{\rm E}}
\newcommand{\thetae}{\theta_{\rm E}}

\newcommand{\pie}{\pi_{\rm E}}
\newcommand{\pien}{\pi_{{\rm E},N}}
\newcommand{\piee}{\pi_{{\rm E},E}}
\newcommand{\dl}{D_{\rm L}}
\newcommand{\ds}{D_{\rm S}}
\newcommand{\hjdp}{{\rm HJD}^\prime}
\newcommand{\hjd}{{\rm HJD}}

\definecolor{brown}{rgb}{0.59, 0.29, 0.0}
\definecolor{darkgreen}{rgb}{0.0, 0.42, 0.24}
\definecolor{darkblue}{rgb}{0.01, 0.31, 0.59}
\definecolor{darkblue}{rgb}{0.0, 0.25, 0.42}
\definecolor{blue}{rgb}{0.0,0.0,1.0}
\definecolor{green}{rgb}{0.0,1.0,0.0}

\begin{document}

\title{Brown dwarf companions in binaries detected from the 2021 season high-cadence microlensing surveys}
\titlerunning{Microlensing brown-dwarf companions in binaries}

\author{
     Cheongho~Han\inst{01}
\and Youn~Kil~Jung\inst{02,03}
\and Ian~A.~Bond\inst{04}
\\
(Leading authors)\\
     Sun-Ju~Chung\inst{02, 05}
\and Michael~D.~Albrow\inst{06}
\and Andrew~Gould\inst{07,08}
\and Kyu-Ha~Hwang\inst{02}
\and Chung-Uk~Lee\inst{02}
\and Yoon-Hyun~Ryu\inst{02}
\and In-Gu~Shin\inst{05}
\and Yossi~Shvartzvald\inst{09}
\and Hongjing~Yang\inst{10}
\and Jennifer~C.~Yee\inst{05}
\and Weicheng~Zang\inst{05,10}
\and Sang-Mok~Cha\inst{02,11}
\and Doeon~Kim\inst{01}
\and Dong-Jin~Kim\inst{02}
\and Seung-Lee~Kim\inst{02}
\and Dong-Joo~Lee\inst{02}
\and Yongseok~Lee\inst{02,11}
\and Byeong-Gon~Park\inst{02}
\and Richard~W.~Pogge\inst{08}
\\
(The KMTNet collaboration)\\
     Fumio~Abe\inst{12}
\and Richard~Barry\inst{13}
\and David~P.~Bennett\inst{13,14}
\and Aparna~Bhattacharya\inst{13,14}
\and Hirosame~Fujii\inst{12}
\and Akihiko~Fukui\inst{15,16}
\and Ryusei~Hamada\inst{17}
\and Yuki~Hirao\inst{18} 
\and Stela~Ishitani Silva\inst{13,19}
\and Yoshitaka~Itow\inst{12}
\and Rintaro~Kirikawa\inst{17}
\and Naoki~Koshimoto\inst{17}
\and Yutaka~Matsubara\inst{12}
\and Shota~Miyazaki\inst{20} 
\and Yasushi~Muraki\inst{12}
\and Greg~Olmschenk\inst{13}
\and Cl{\'e}ment~Ranc\inst{21}
\and Nicholas~J.~Rattenbury\inst{22}
\and Yuki~Satoh\inst{17}
\and Takahiro~Sumi\inst{17}
\and Daisuke~Suzuki\inst{17}
\and Mio~Tomoyoshi\inst{17}
\and Paul~J.~Tristram\inst{23}
\and Aikaterini~Vandorou\inst{13,14}
\and Hibiki~Yama\inst{17}
\and Kansuke~Yamashita\inst{17}
\\
(The MOA Collaboration)\\
}

\institute{
      Department of Physics, Chungbuk National University, Cheongju 28644, Republic of Korea,                                                          
\and  Korea Astronomy and Space Science Institute, Daejon 34055, Republic of Korea                                                                     
\and  Korea University of Science and Technology (UST), 217 Gajeong-ro, Yuseong-gu, Daejeon, 34113, Republic of Korea                                  
\and  Institute of Natural and Mathematical Science, Massey University, Auckland 0745, New Zealand                                                     
\and  Center for Astrophysics $|$ Harvard \& Smithsonian, 60 Garden St., Cambridge, MA 02138, USA                                                      
\and  University of Canterbury, Department of Physics and Astronomy, Private Bag 4800, Christchurch 8020, New Zealand                                  
\and  Max-Planck-Institute for Astronomy, K\"{o}nigstuhl 17, 69117 Heidelberg, Germany                                                                 
\and  Department of Astronomy, Ohio State University, 140 W. 18th Ave., Columbus, OH 43210, USA                                                        
\and  Department of Particle Physics and Astrophysics, Weizmann Institute of Science, Rehovot 76100, Israel                                            
\and  Department of Astronomy, Tsinghua University, Beijing 100084, China                                                                              
\and  School of Space Research, Kyung Hee University, Yongin, Kyeonggi 17104, Republic of Korea                                                        
\and  Institute for Space-Earth Environmental Research, Nagoya University, Nagoya 464-8601, Japan                                                      
\and  Code 667, NASA Goddard Space Flight Center, Greenbelt, MD 20771, USA                                                                             
\and  Department of Astronomy, University of Maryland, College Park, MD 20742, USA                                                                     
\and  Komaba Institute for Science, The University of Tokyo, 3-8-1 Komaba, Meguro, Tokyo 153-8902, Japan                                               
\and  Instituto de Astrof{\'i}sica de Canarias, V{\'i}a L{\'a}ctea s/n, E-38205 La Laguna, Tenerife, Spain                                             
\and  Department of Earth and Space Science, Graduate School of Science, Osaka University, Toyonaka, Osaka 560-0043, Japan                             
\and  Institute of Astronomy, Graduate School of Science, The University of Tokyo, 2-21-1 Osawa, Mitaka, Tokyo 181-0015, Japan                         
\and  Department of Physics, The Catholic University of America, Washington, DC 20064, USA                                                             
\and  Institute of Space and Astronautical Science, Japan Aerospace Exploration Agency, 3-1-1 Yoshinodai, Chuo, Sagamihara, Kanagawa 252-5210, Japan   
\and  Sorbonne Universit\'e, CNRS, UMR 7095, Institut d'Astrophysique de Paris, 98 bis bd Arago, 75014 Paris, France                                   
\and  Department of Physics, University of Auckland, Private Bag 92019, Auckland, New Zealand                                                          
\and  University of Canterbury Mt.~John Observatory, P.O. Box 56, Lake Tekapo 8770, New Zealand                                                        
}


\abstract
{}
{
As a part of the project aiming to build a homogeneous sample of binary-lens (2L1S) events 
containing brown-dwarf (BD) companions,  we investigate the 2021 season microlensing data 
collected by the Korea Microlensing Telescope Network (KMTNet) survey.
}
{
For this purpose, we first identify 2L1S events by conducting systematic analyses of 
anomalous lensing events.  We then select candidate BD-companion events by applying the 
criterion that the mass ratio between the lens components is less than $q_{\rm th}\sim 0.1$. 
}
{
From this procedure, we find four binary-lens events including KMT-2021-BLG-0588, 
KMT-2021-BLG-1110, KMT-2021-BLG-1643, and KMT-2021-BLG-1770, for which the estimated mass 
ratios are $q\sim 0.10$, 0.07, 0.08, and 0.15, respectively. The event KMT-2021-BLG-1770 
is selected as a candidate despite the fact that the mass ratio is slightly greater than 
$q_{\rm th}$ because the lens mass expected from the measured short time scale of the event, 
$\te\sim 7.6$~days, is small.  From the Bayesian analyses, we estimate that the primary and 
companion masses are
$(M_1/M_\odot, M_2/M_\odot)= 
 (0.54^{+0.31}_{-0.24}, 0.053^{+0.031}_{-0.023})$ for KMT-2021-BLG-0588L, 
$(0.74^{+0.27}_{-0.35}, 0.055^{+0.020}_{-0.026})$ for KMT-2021-BLG-1110L, 
$(0.73^{+0.24}_{-0.17}, 0.061^{+0.020}_{-0.014})$ for KMT-2021-BLG-1643L, and 
$(0.13^{+0.18}_{-0.07}, 0.020^{+0.028}_{-0.011})$ for KMT-2021-BLG-1770L.  
It is estimated that the probabilities of the lens companions being in the BD mass range are 
82\%, 85\%, 91\%, and 59\% for the individual events.  For confirming the BD nature of the 
lens companions found in this and previous works by directly imaging the lenses from future 
high-resolution adaptive-optics (AO) followup observations, we provide the lens-source 
separations expected in 2030, which is an approximate year of the first AO light on 30~m class 
telescopes.
}
{}

\keywords{brown dwarfs -- gravitational lensing: micro}

\maketitle

\section{Introduction}\label{sec:one}

With the trait that does not depend on the light of a lens, microlensing is suited for finding 
and studying faint and dark astronomical objects. One scientifically important object to which 
this microlensing trait is successfully applied is an extrasolar planet. With the proposals of 
\citet{Mao1991} and \citet{Gould1992b}, extensive searches for extrasolar planets using the
microlensing method have been carried out since the 1990s. Being started with the first 
discovery of a giant planet in 2003 by \citet{Bond2004}, 200 microlensing planets have been 
reported according to the NASA Exoplanet Archive\footnote{https://exoplanetarchive.ipac.caltech.edu}, 
making microlensing the method that is used to detect the third most planets after the transit and 
radial-velocity methods.

Brown dwarfs (BDs) are another population of astronomical objects for which microlensing is
well suited for detections.
Microlensing BDs can be detected through two channels.  The first channel is 
via a single-lens single-source (1L1S) event with a short time scale $\te$.  The event time 
scale is related to the lens mass $M$ as 
\begin{equation}
\te = {\thetae\over \mu}; \qquad
\thetae = (\kappa M \pi_{\rm rel})^{1/2}, 
\label{eq1}
\end{equation}
and thus short time-scale events may be produced by BDs with masses lower than those of stars.
Here $\thetae$ represents the angular Einstein radius, $\mu$ is the relative lens-source proper 
motion, $\kappa=4G/(c^2{\rm AU})$, $\pi_{\rm rel} = {\rm AU}(D_{\rm L}^{-1}- D_{\rm S}^{-1})$ 
is the relative lens-source parallax, and $\dl$ and $\ds$ denote the distances to the lens and 
source, respectively. However, it is difficult to confirm the BD nature of a lens based on the 
event time scale alone, because the time scale depends additionally on $\mu$ and $\pi_{\rm rel}$. 
The mass and distance to the lens can be unambiguously determined by measuring the extra observables 
of the Einstein radius $\thetae$ and the microlens parallax $\pie$ from the relations 
\begin{equation}
M= {\thetae \over \kappa\pie};\qquad
\dl = {{\rm AU} \over \pie\thetae + \pi_{\rm S}}.
\label{eq2}
\end{equation}
The microlens parallax is related to the relative lens-source parallax and Einstein radius 
by $\pie =\pi_{\rm rel}/\thetae$ \citep{Gould1992a, Gould2000}. For a 1L1S event, the probability 
of measuring the angular Einstein radius is very low because $\thetae$ can be measured for only a very 
minor fraction of events in which the lens passes over the surface of the source, for example, 1L1S 
events presented in \citet{Han2020}, \citet{Gould2022b}, and \citet{Koshimoto2023}.  The probability 
of measuring the microlens parallax, which is generally measured from the deviation of the lensing 
light curve caused by the departure of the relative lens-source motion from rectilinear induced by 
the orbital motion of Earth, is even lower because the parallax-induced deviation in the lensing 
light curve is generally too small to be measured for a short time-scale BD event.  The microlens 
parallax for a short time-scale event can be measured under special observational environments, 
and there exist only three cases for which the nature of the single BD lens was confirmed from the 
mass determination by measuring the microlens parallax.  The first case is OGLE-2007-BLG-224, for 
which $\pie$ was measured from the subtle differences among the light curves constructed from 
observations using telescopes lying at multiple sites on Earth when the magnifications of the event 
were extremely high \citep{Gould2009}. For the other two cases of OGLE-2015-BLG-1268 \citep{Zhu2016} 
and OGLE-2017-BLG-0896 \citep{Shvartzvald2019}, $\pie$ values were measured from simultaneous 
observations of the events using ground-based telescopes and the space-based Spitzer satellite. 
In the case of OGLE-2015-BLG-1482 \citep{Chung2017}, which was also simultaneously observed using 
the Spitzer and ground-based telescopes, the light curve was almost equally well explained by two 
solutions, in which the lens was a very low-mass star with a mass $0.10 \pm 0.02~M_\odot$ according 
to one solution, and the lens is a BD with a mass $0.052 \pm 0.008~M_{\rm J}$ according to the other 
solution, and thus the BD nature of the lens could not be confirmed.

Another channel of detecting microlensing BDs is via a binary-lens single-source (2L1S) event.
Compared to a 1L1S event, analysis of a 2L1S event yields an additional constraint of the
companion-to-primary mass ratio $q$. This constraint can be used to select candidate BD
companions of binary lenses based on the fact that typical Galactic lensing events are produced 
by low-mass stars \citep{Han2003}, and thus companions with mass ratios $q \lesssim 0.1$ are very
likely to be BDs. Furthermore, the probability of measuring the Einstein radii for these events is
high because the light curves of these events usually exhibit anomaly features resulting from source 
crossings over or approaches very close to caustics.  In these cases, the light curves are likely to 
be affected by finite-source effects, from which $\thetae$ can be measured and the lens mass can be 
further constrained.

\begin{table*}[t]
\footnotesize
\caption{Source positions and extinction.\label{table:one}}
\begin{tabular}{lccc}
\hline\hline
\multicolumn{1}{c}{Event}                     &
\multicolumn{1}{c}{(RA, DEC)$_{\rm J2000}$}   &
\multicolumn{1}{c}{$(l, b)$}                  &
\multicolumn{1}{c}{$A_I$}                     \\
\hline
KMT-2021-BLG-0588     &  (18:06:53.32, -27:25:19.31)  &  (3.666, -3.291)    &  0.92    \\
KMT-2021-BLG-1110     &  (17:55:18.86, -30:22:24.49)  &  (-0.156, -2.545)   &  2.10    \\
KMT-2021-BLG-1643     &  (18:02:32.53, -30:36:18.68)  &  (0.417, -4.012)    &  1.30    \\
KMT-2021-BLG-1770     &  (18:01:45.59, -27:25:07.21)  &  (3.112, -2.296)    &  1.86    \\
\hline                                                 
\end{tabular}
\end{table*}

In order to find BDs through the second channel, \citet{Han2022}, hereafter paper~I, investigated 
the microlensing data collected during the 2016--2018 period by the high-cadence surveys and 
reported 6 binaries with candidate BD companions, including OGLE-2016-BLG-0890LB, MOA-2017-BLG-477LB, 
OGLE-2017-BLG-0614LB, KMT-2018-BLG-0357LB, OGLE-2018-BLG-1489LB, and OGLE-2018-BLG-0360LB. From 
continued analyses of the lensing events found during the 2018--2020 period, \citet{Han2023}, 
hereafter paper~II, reported another 4 binaries with candidates BD companions, including 
KMT-2018-BLG-0321LB, KMT-2018-BLG-0885LB, KMT-2019-BLG-0297LB, and KMT-2019-BLG-0335LB.

In this work, we report four additional candidate BD companions to binary lenses found from the 
inspection of the 2021 season microlensing data, including KMT-2021-BLG-0588LB, KMT-2021-BLG-1110LB, 
KMT-2021-BLG-1643LB, and KMT-2021-BLG-1770LB.  The main scientific  purpose of this and previous 
works is building a homogeneous sample of binary-lens events containing BD companions found from 
the KMTNet survey by applying a consistent criterion.  The sample will be useful for future statistical 
analyses on BDs such as the distribution of mass ratios and separations and the occurrence rate of 
star-BD binary pairs.

For the presentation of the findings and analyses of the BD events, we organize the paper as
follows. In Sect.~\ref{sec:two}, we describe the procedure of selecting candidate events produced 
by binary lenses possessing BD companions.  In Sect.~\ref{sec:three}, we depict the data used in 
the analyses and the observations carried out to obtain the data. In Sect.~\ref{sec:four}, we start 
by explaining the common procedure applied to analyze the events and detail the analyses of the 
individual events in the following subsections: KMT-2021-BLG-0588L in Sect.~\ref{sec:four-one}, 
KMT-2021-BLG-1110L in Sect.~\ref{sec:four-two}, KMT-2021-BLG-1643L in Sect.~\ref{sec:four-three}, 
and KMT-2021-BLG-1770L in Sect.~\ref{sec:four-four}. In Sect.~\ref{sec:five}, we mention the 
procedure of specifying the source stars and estimate the Einstein radii of the individual events. 
In Sect.~\ref{sec:six}, we explain the Bayesian analyses conducted to estimate the physical lens 
parameters of the events and present the obtained parameters. In Sect.~\ref{sec:seven}, we summarize 
the results from the analyses and discuss future followup observations that can confirm the BD 
natures of the lens companions reported in this work and those found from previous analyses in 
papers~I and II.

\section{Selections of BD candidates }\label{sec:two}

The binary-lens events with BD companions were found from the inspection of the microlensing 
events that were found in the 2021 season by the Korea Microlensing Telescope Network 
\citep[KMTNet:][]{Kim2016} survey. For a 2L1S event possessing a planetary lens companion, 
with a companion-to-primary mass ratio of order $10^{-3}$ or less, the signal of the companion, 
in general, can be readily identified from its characteristic short-term anomaly feature in the 
lensing light curve \citep{Gould1992b}. For a 2L1S event with a BD companion, which has a mass 
ratio of order $10^{-2}$, however, it is difficult to promptly identify the BD nature of the 
companion, because the lensing light curves are, in many cases, similar to those produced by 
binary lenses with approximately equal-mass components.  In the searches for BD companions in 
binary lenses, therefore, we conducted systematic analyses of all anomalous lensing events 
detected by the KMTNet survey.

We selected events with BD companions by imposing the criterion of $q \lesssim 0.1$ among the 
2L1S events identified from the first-round analyses. We note that the criterion is the same as 
the criterion that was adopted in papers I and II, and thus the BD events presented in this and 
previous works constitute a uniform sample. From this procedure, we identified four candidate 
BD-companion events including KMT-2021-BLG-0588, KMT-2021-BLG-1110, KMT-2021-BLG-1643, and 
KMT-2021-BLG-1770.  In Table~\ref{table:one}, we list the equatorial coordinates, 
(RA, DEC)$_{\rm J2000}$, of the individual events together with the corresponding Galactic 
coordinates, $(l, b)$, and $I$-band extinction, $A_I$, toward the field. Here the extinction 
values were adopted from the OGLE Internet archive \citep{Nataf2013}.\footnote{ \text 
http://ftp.astrouw.edu.pl/ogle/ogle3/ext/blg/} The event KMT-2021-BLG-1770 was picked out 
despite the fact that the estimated mass ratio between the lens components, $q\sim 0.15$, was 
slightly greater than the adopted threshold mass ratio $q_{\rm th}\sim 0.1$, because the mass 
of the lens expected from the short time scale of the event, $\te\sim 7.6$~days, was low, and 
thus the probability for the mass of the companion to be in the BD mass regime was high.  
For this reason, this event is not a part of uniformly selected sample for future statistical 
studies, although analysis is presented in this work.  For 
the identified candidate events, we then checked whether the events were additionally observed 
by other lensing surveys to include the data in the analyses if they exist. We found that 
KMT-2021-BLG-0588 was additionally observed by the Microlensing Observations in Astrophysics 
\citep[MOA:][]{Bond2001} group, who referred to the event as MOA-2021-BLG-139, and the other 
events were observed solely by the KMTNet group. For KMT-2021-BLG-0588, we use the KMTNet ID 
reference because the KMTNet group first found the event.

\section{Observations and data}\label{sec:three}

The KMTNet group has carried out a high-cadence survey since 2016 by monitoring stars lying
toward the Galactic bulge field in search of light variation of stars caused by microlensing. 
The survey group utilizes three wide-field telescopes, which are distributed in three sites of 
the Southern Hemisphere for continuous and dense coverage of lensing events. The sites of the 
individual telescopes are the Siding Spring Observatory in Australia (KMTA), the Cerro Tololo 
interamerican Observatory in Chile (KMTC), and the South African Astronomical Observatory in 
South Africa (KMTS). The telescopes are identical and each telescope with a 1.6~m aperture is 
equipped with a camera that yields 4~deg$^2$ field of view. KMTNet observations were mainly 
conducted in the $I$ band, which is relatively less affected by extinction, and about one tenth 
of images were acquired in the $V$-band for the source color measurements of lensing events. 
Photometry of the events was conducted using the automatized pySIS pipeline \citep{Albrow2009}, 
which is based on the difference image method \citep{Tomaney1996, Alard1998}. For the color 
measurements of the source stars, we additionally used the pyDIA code \citep{Albrow2017} to 
construct a set of the $I$ and $V$-band light curves and color-magnitude diagrams (CMDs) of 
stars that lie in the neighborhoods of the source stars.  For the events analyzed in this work, 
we conducted rereduction of the data to obtain optimized photometry data after the events were 
selected as BD candidates.  We normalized the error bars of the data to make them consistent 
with scatter of data and $\chi^2$ per degree of freedom (dof) for each data set to become unity. 
In the error-bar normalization process, we used the routine described in \citet{Yee2012}.

Among the four analyzed events, the lensing event KMT-2021-BLG-0588 was additionally observed 
by the MOA survey. The observations of the event by the MOA survey were done with the use of 
the 1.8~m telescope of the Mt.~John Observatory in New Zealand. The camera mounted on the 
telescope yields 2.2~deg$^2$ field of view.  The MOA observations were mostly conducted in 
the customized MOA-$R$ band, and the photometry was done using the MOA pipeline. Normalization 
of the MOA data set was done using the same routine that was applied to the KMTNet data sets. 
\footnote{The photometry data are available at the follow site: \\
http://astroph.chungbuk.ac.kr/$\sim$cheongho/download.html.}

\section{Analyses}\label{sec:four}

The events were analyzed under the common interpretation of the lens-system configuration that 
the lenses are binaries because the light curves of all events exhibit caustic features that 
arise due to the multiplicity of the lens masses. Under the assumption of a rectilinear relative 
lens-source motion, the lensing light curve of a 2L1S event is described by 7 basic lensing 
parameters. Among these parameters, the first three parameters $(t_0, u_0, \te)$ describe the 
lens-source approach, and the individual parameters represent the time of the closest lens-source 
approach, the lens-source separation at $t_0$, and the event time scale, respectively. Another 
three parameters $(s, q, \alpha)$ describe the binarity of the lens, and the individual parameters 
describe the projected separation (scaled to $\thetae$) and mass ratio between the lens components, 
and the angle between the source trajectory and the axis connecting the binary lens components. 
The last parameter $\rho$ represents the ratio of the angular source radius $\theta_*$ to the 
Einstein radius, $\rho=\theta_*/\thetae$ (normalized source radius), and it describes the 
deformation of the light curve during the caustic crossings of a source caused by finite-source 
effects.

A 2L1S lensing light curve can deviate from a standard form due to the departure of the relative
lens-source motion from rectilinear. The first cause of such a deviation is the microlens-parallax
effects, which is caused by the positional change of the observer by the orbital motion of Earth
around the sun \citep{Gould1992a}. The second cause is the lens-orbital effects, which is caused 
by the change of the lens position by the orbital motion of the binary lens \citep{Dominik1998}. 
These higher-order effects induce subtle deviations in the lensing light curve from the standard 
form, and description of these deviations requires additional lensing parameters in modeling.  We 
checked these higher-order effects by conducting additional modeling, in which additional parameters 
were added in the modeling.  The two parameters describing the parallax effect are $(\pien, \piee)$, 
which represent the north and east components of the microlens-lens parallax vector $\pivec_{\rm E} 
= (\pi_{\rm rel}/ \thetae)(\muvec/\mu)$, respectively.  Under the assumption that the positional 
change of the lens by the orbital motion is minor, the lens-orbital effect is described by two 
parameters $(ds/dt, d\alpha/dt)$, which denote the annual change rate of the binary separation 
and source trajectory angle, respectively.  It was found that secure detections of the higher-order 
effects were difficult for KMT-2021-BLG-0588, KMT-2021-BLG-1110, and KMT-2021-BLG-1770, for which 
the event time scales are less than 40~days.  For KMT-2021-BLG-1643 with $\te\sim 105$~days, the 
higher-order effects are minor, but the amplitude of the parallax parameters yielded a useful 
constraint on the physical lens parameters.  See Sect.~\ref{sec:six} for the detailed discussion 
on the parallax constraint.

In the 2L1S modeling, we searched for a lensing solution, which refers to a set of the lensing
parameters that best depict the observed lensing light curve. In the first round of modeling, 
we divided the lensing parameters into two groups, and found the binary parameters $(s, q)$ of 
the first group via a grid approach with multiple initial values of $\alpha$, and the other lensing 
parameters of the second group were searched for by minimizing $\chi^2$ using the Markov Chain Monte 
Carlo (MCMC) method with an adaptive step size Gaussian sampler \citep{Doran2004}.  In the second 
round, we refined the local solutions identified from the first-round modeling by further reducing 
$\chi^2$ value using the MCMC method.  We adopt this two-step approach because the change of the 
lensing magnification with the variation of the grid parameters is discontinuous, while the 
magnification changes smoothly with the variation of the downhill parameters.  Furthermore, the 
$\Delta\chi^2$ map obtained from the first-round grid search enables us to identify local solutions 
that are caused by various types of degeneracy.  We consider the limb-darkening variation of the 
source surface brightness in the computation of finite magnifications by adopting the linear 
limb-darkening coefficients of \citet{Claret2000} corresponding to the stellar type of the source 
stars.  In the following subsections, we present the detailed analyses conducted for the individual 
events.

\subsection{KMT-2021-BLG-0588}\label{sec:four-one}

\begin{figure}[t]
\includegraphics[width=\columnwidth]{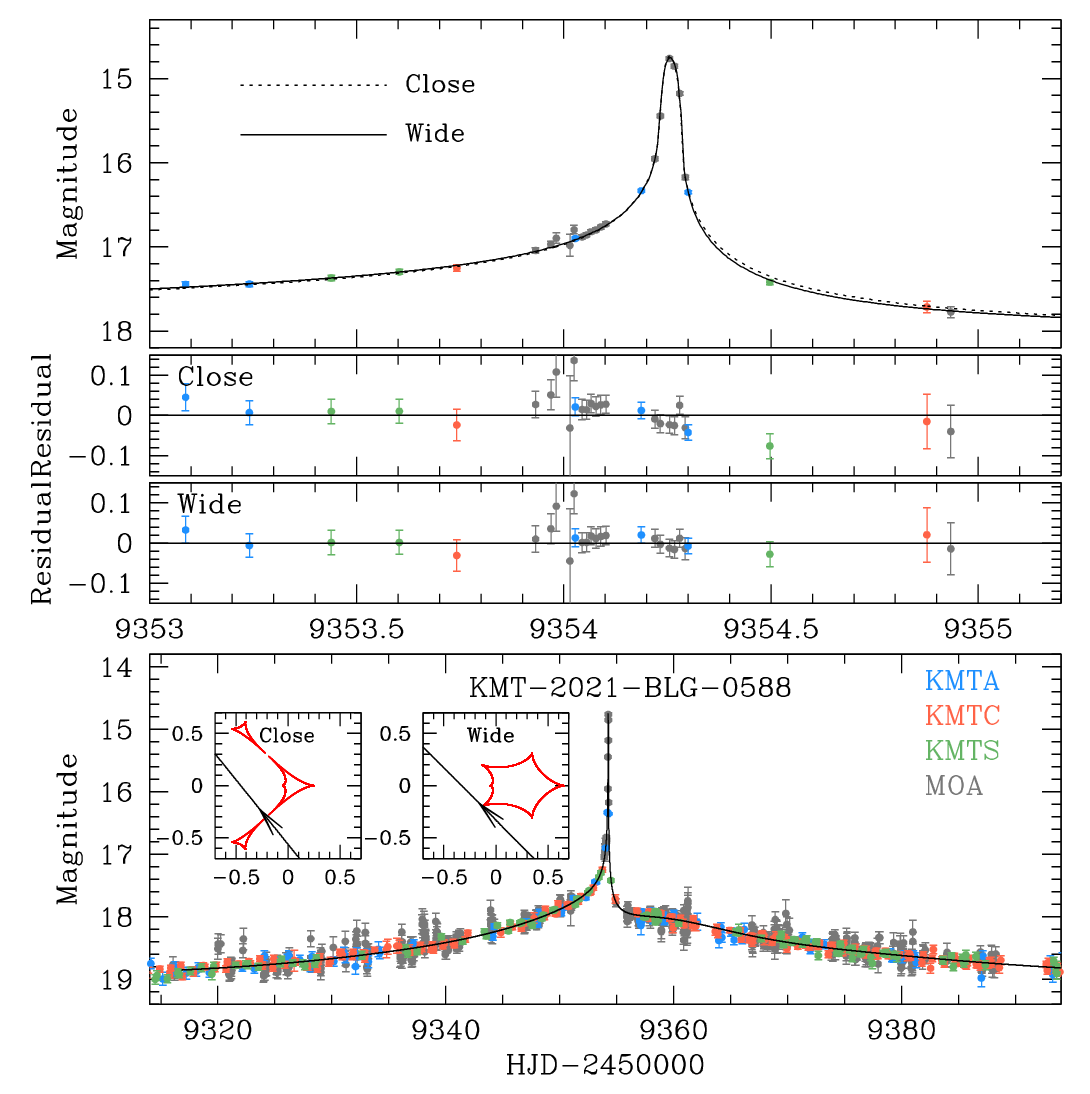}
\caption{
Light curve of KMT-2021-BLG-0588. The bottom panel shows the whole view of the light curve, 
and the upper panels show the zoom-in view of the light curve, models, and residuals in the 
region around the strong peak. The dotted and solid curves drawn over the data points are the 
models of the close and wide solutions, respectively. The two insets in the bottom panel show 
the lens-system configurations of the close and wide models. 
}
\label{fig:one}
\end{figure}

Figure~\ref{fig:one} shows the lensing light curve of the event KMT-2021-BLG-0588. The source 
with an $I$-band baseline magnitude $I_{\rm base}\sim 19.11$ was in the KMT32 field, toward 
which observations were conducted with a 2.5~hr cadence. The source flux magnification induced 
by lensing was first found by the KMTNet group on 2021 April 26, which corresponds to the abridged 
heliocentric Julian date $\hjdp\equiv \hjd - 2450000 =9331$, when the source was brighter than the 
baseline by $\Delta I\sim 0.46$~mag.  The light curve exhibited a strong anomaly, which peaked at 
$\hjdp\sim 9354.25$ with a strong deviation of $\Delta I\sim 3$~mag from the baseline 1L1S model.  
The MOA group independently found the event on 2021 May 22 ($\hjdp=9357$), which was about 3 days 
after the strong peak.  The zoom-in view of the strong peak, which was covered by the combination 
of the MOA and KMTA data sets, is shown in the top panel of Figure~\ref{fig:one}.  From the sharp 
rise and fall, the strong peak is likely to be produced by the source star's crossing over the tip 
of a caustic formed by a binary lens.

In Table~\ref{table:two}, we list the lensing parameters of the solutions found from the 2L1S 
modeling of the light curve together with the $\chi^2$ values of the fits and degrees of freedom 
(dof). We identified a pair of local solutions, in which one solution has a binary separation 
$s < 1$ (close solution) and the other solution has a separation $s > 1$ (wide solution). Although 
the solutions are designated as the "close" and "wide" solutions, we note that the similarity 
between the model curves of the two solutions is caused by an accidental degeneracy rather than 
the well-known close--wide degeneracy, which arises due to the similarity between the central 
caustics induced by a pair of solutions with separations $s$ and $1/s$ \citep{Griest1998, 
Dominik1999, An2005}. We further discuss the cause of the degeneracy in the following paragraph. 
It is found that the wide solution with $s\sim 1.17$ yields a better fit than the close solution 
with $s\sim 0.77$ by $\Delta\chi^2=71.8$, and thus the degeneracy is resolved with strong 
statistical confidence.

\begin{table}[t]
\small
\caption{Model parameters of KMT-2021-BLG-0588.\label{table:two}}
\begin{tabular*}{\columnwidth}{@{\extracolsep{\fill}}lcccc}
\hline\hline
\multicolumn{1}{c}{Parameter}   &
\multicolumn{1}{c}{Close}       &
\multicolumn{1}{c}{Wide}        \\
\hline
$\chi^2$/dof            &  $1748.2/1673        $    &  $1676.4/1673         $     \\
$t_0$ (HJD$^\prime$)    &  $9357.453 \pm 0.069 $    &  $9356.157 \pm 0.065  $     \\
$u_0$                   &  $0.364 \pm 0.015    $    &  $0.233 \pm 0.008     $     \\
$\te$ (days)            &  $29.22 \pm 0.89     $    &  $39.34 \pm 1.24      $     \\
$s$                     &  $0.7649 \pm 0.0014  $    &  $1.1717 \pm 0.0077   $     \\
$q$                     &  $0.1093 \pm 0.0024  $    &  $0.0992 \pm 0.0038   $     \\
$\alpha$ (rad)          &  $0.8979 \pm 0.0095  $    &  $0.7883 \pm 0.0092   $     \\
$\rho$ ($10^{-3}$)      &  $0.973 \pm 0.032    $    &  $0.700 \pm 0.024     $     \\
\hline
\end{tabular*}
\tablefoot{ ${\rm HJD}^\prime = {\rm HJD}- 2450000$.  }
\end{table}

In Figure~\ref{fig:one}, we draw the model curve of the wide solution in the bottom panel, which 
shows the whole view of the light curve, and plot the models curves and residuals of both the 
close and wide solutions in the upper panels, which show the zoom-in view of the region around 
the strong peak. According to the wide solution, the estimated event time scale and the mass 
ratio between the lens components are $\te \sim 39$~days and $q\sim 0.10$, respectively. From 
the fact that the time scale is in the range of events produced by stellar lenses together with 
the fact that the mass ratio is low, the probability of the binary lens companion being a BD is 
high. The normalized source radius, $\rho\sim 0.7\times 10^{-3}$, was securely measured from the 
analysis of the strong peak, which was affected by finite-source effects

The lens-system configurations of the close and wide solutions are presented in the two insets 
of the bottom panel of Figure~\ref{fig:one}. According to the wide solution, the binary lens 
forms a single six-sided resonant caustic, and the strong peak was produced by the source 
passage through the tip of the lower left cusp of the caustic. According to the close solution, 
on the other hand, the lens induces 3 sets of caustics, in which a single central caustic around 
the primary lens is detached from the two peripheral caustics, and the strong peak was generated 
by the source crossing over the slim cusp extending from the lower left cusp of the central 
caustic. The two sets of caustics of the close and wide solutions do not appear to be similar 
to each other, and this suggests that the degeneracy between the two solutions is accidental.

\subsection{KMT-2021-BLG-1110}\label{sec:four-two}

\begin{figure}[t]
\includegraphics[width=\columnwidth]{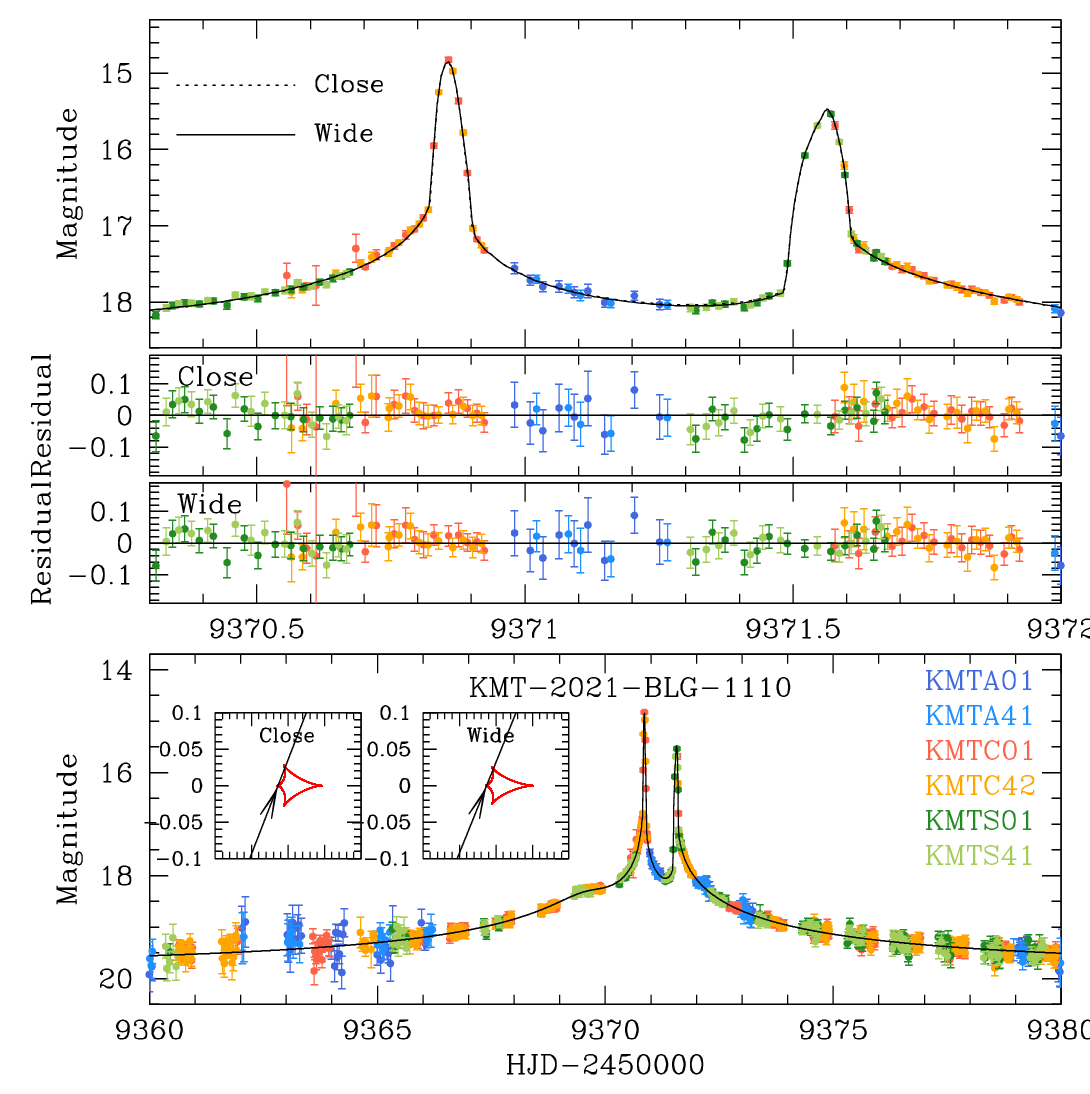}
\caption{
Lensing light curve of KMT-2021-BLG-1110.  The layout and scheme of the plot are same as 
those in Fig.~\ref{fig:one}. 
}
\label{fig:two}
\end{figure}

We present the light curve of the lensing event KMT-2021-BLG-1110 in Figure~\ref{fig:two}. The 
lensing magnification of the source, which had a baseline magnitude $I_{\rm base}\sim 19.52$ 
before lensing, was found by the KMTNet group on 2021 June 2 ($\hjdp=9367$), when the source was 
brighter than the baseline by $\Delta I\sim 0.5$~mag.  The source lies in the overlapping region 
of the KMTNet prime fields BLG01 and BLG41, toward which observations were done with a 0.5~hr 
cadence for each field, and a 0.25~hr cadence in combination.  The light curve is characterized 
by the double spikes appearing at $t_1\sim 9370.85$ and $t_2\sim 9371.56$. The rising and falling 
sides of both spikes were densely and continuously resolved from the high-cadence observations 
conducted with the use of the three KMTNet telescopes.  The first spike was resolved by the KMTC 
data, and the second one was covered by the combined data from KMTS and KMTC.

\begin{table}[t]
\small
\caption{Model parameters of KMT-2021-BLG-1110.\label{table:three}}
\begin{tabular*}{\columnwidth}{@{\extracolsep{\fill}}lcccc}
\hline\hline
\multicolumn{1}{c}{Parameter}   &
\multicolumn{1}{c}{Close}       &
\multicolumn{1}{c}{Wide}        \\
\hline
$\chi^2$/dof            &  $6113.9/6108           $    &  $6080.10/6108         $     \\
$t_0$ (HJD$^\prime$)    &  $9370.9888 \pm 0.0023  $    &  $9370.9974 \pm 0.0022 $     \\
$u_0$ (10$^{-2}$)       &  $1.272 \pm 0.025       $    &  $1.176 \pm 0.032      $     \\
$\te$ (days)            &  $26.97 \pm 0.53        $    &  $29.82 \pm 0.78       $     \\
$s$                     &  $0.4376 \pm 0.0031     $    &  $2.4326 \pm 0.0192    $     \\
$q$                     &  $0.0726 \pm 0.0018     $    &  $0.0742 \pm 0.0024    $     \\
$\alpha$ (rad)          &  $1.9408 \pm 0.0048     $    &  $1.9541 \pm 0.0047    $     \\
$\rho$ ($10^{-3}$)      &  $0.886 \pm 0.024       $    &  $0.793 \pm 0.025      $     \\
\hline
\end{tabular*}
\end{table}

The spike features are very likely to be produced by the caustic crossings of the source, 
and thus we conducted modeling the light curve under the 2L1S interpretation. The modeling 
yielded two local solutions: one with $s<1$ (close solution) and the other with $s>1$ (wide 
solution).  It is found that the wide solution is preferred over the close solution by 
$\Delta\chi^2 =33.8$, which is large enough to resolve the degeneracy between the solutions. 
The model curve of the wide solution is drawn in the bottom panel of Figure~\ref{fig:two}, and 
the model curves and residuals of both the close and wide solutions in the region around the 
two peaks are presented in the upper panels. The similarity between the models of the two 
solutions is caused by the classic close--wide degeneracy. The lensing parameters of the 
solutions are listed in Table~\ref{table:three} together with the values of $\chi^2$/dof. The 
binary lensing parameters are $(s, q)_{\rm close}\sim (0.44, 0.07)$ for the close solution, 
and $(s, q)_{\rm wide}\sim (2.43, 0.07)$ for the wide solution. From the fact that the estimated 
mass ratio $q\sim 0.07$ between the lens components is low together with the fact that the event 
time scale $\te \sim 27$--29~days is a typical value of a stellar lens event, the companion of 
the lens is a strong BD candidate. The normalized source radius, $\rho\sim 0.79\times 10^{-3}$ 
for the wide solution, is precisely measured from the well-resolved spike features.

\begin{figure}[t]
\includegraphics[width=\columnwidth]{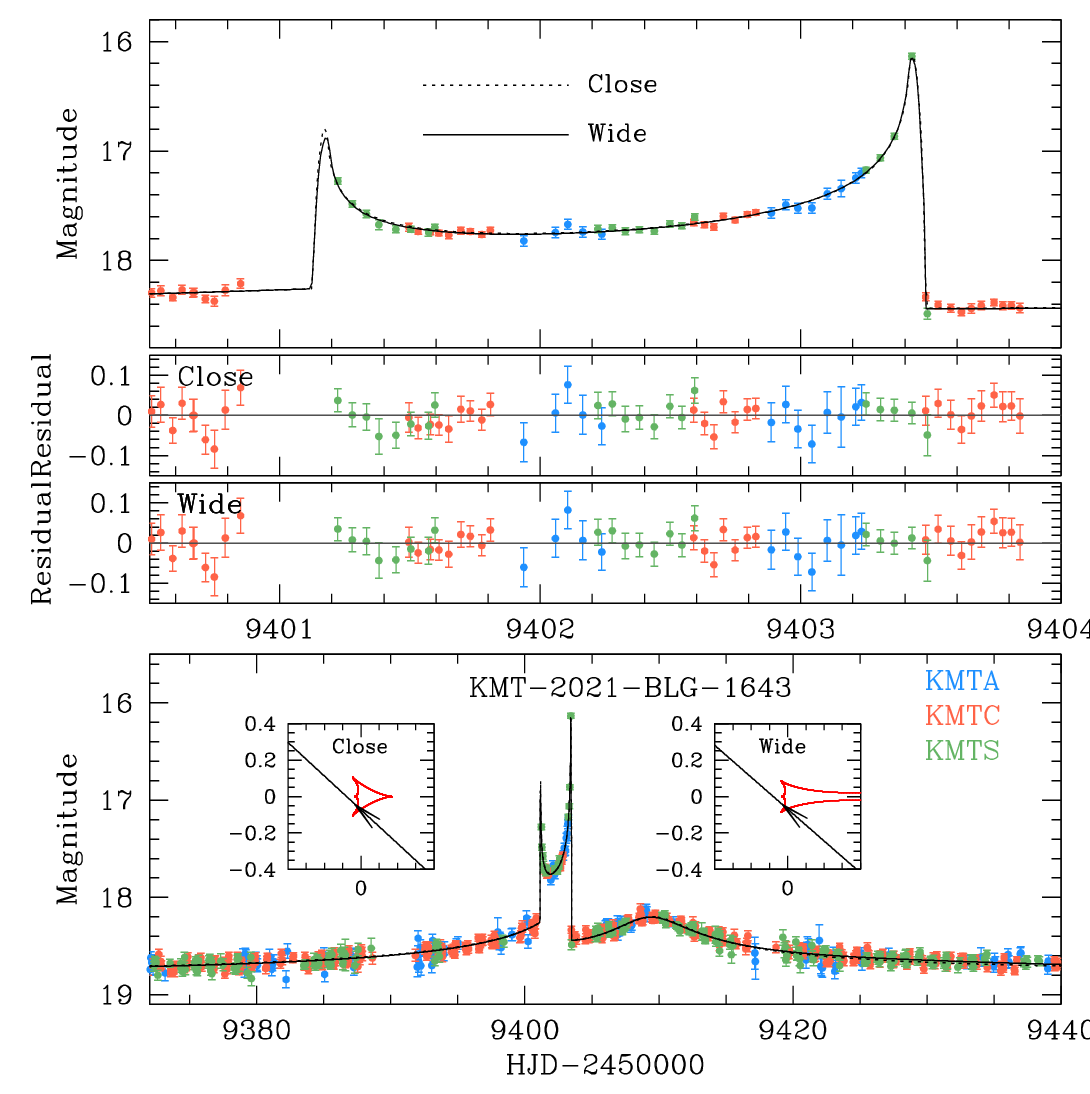}
\caption{
Light curve of the lensing event KMT-2021-BLG-1643. The plot scheme and layout are same as 
those in Fig.~\ref{fig:one}.
}
\label{fig:three}
\end{figure}

In the two insets of the bottom panels of Figure~\ref{fig:two}, we present the lens-system 
configurations of the close and wide solutions. Both solutions result in central caustics of 
similar shape, in which the caustic is elongated along the binary-lens axis. The source passed 
through the back-end side of the caustic at an acute source trajectory angle of $\sim 69^\circ$ 
with respect to the binary axis.  According to the model, the two spikes were produced by the 
successive passages of the source through the on-axis cusp and upper off-axis cusp of the caustic.

\begin{figure}[t]
\includegraphics[width=\columnwidth]{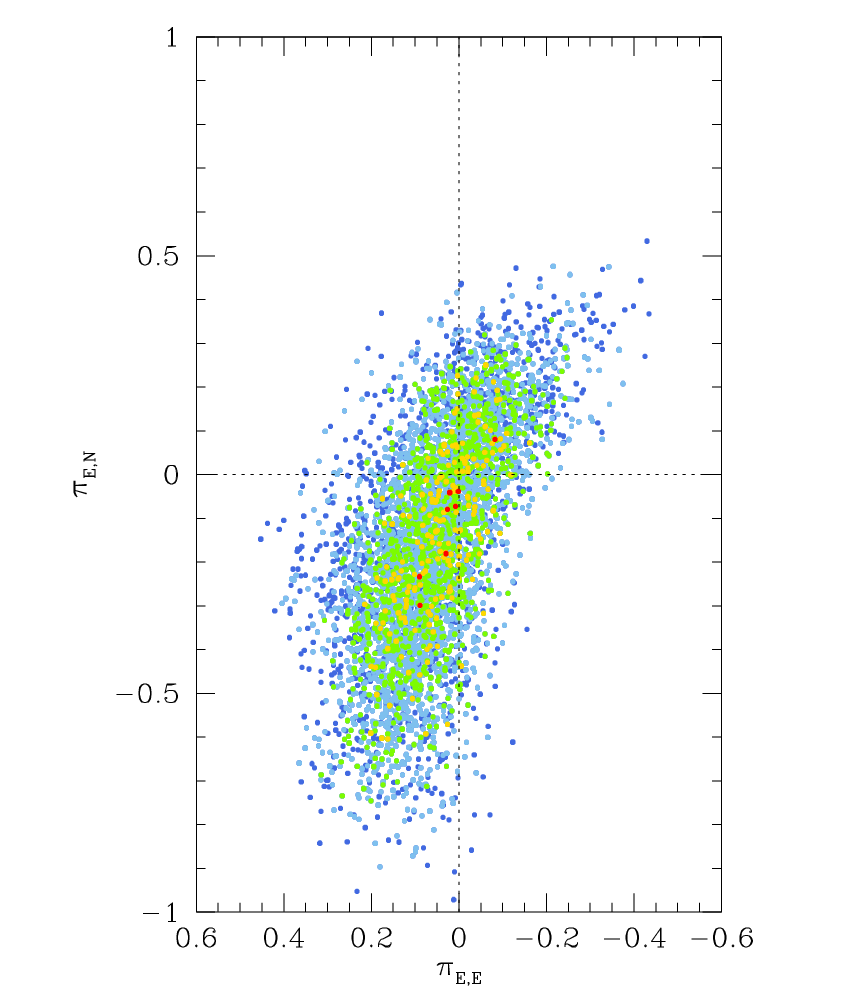}
\caption{
Scatter plot of points in the MCMC chain on the $\piee$--$\pien$ parameter plane 
obtained from the modeling of KMT-2021-BLG-1643 considering higher-order effects.
The color coding is set to represent points with
$\Delta\chi^2\leq 1$  (red), 
$\leq 4$  (yellow), 
$\leq 9$  (green), 
$\leq 16$ (cyan), and
$\leq 25$ (blue). 
}
\label{fig:four}
\end{figure}

\begin{table}[t]
\small
\caption{Model parameters of KMT-2021-BLG-1643.\label{table:four}}
\begin{tabular*}{\columnwidth}{@{\extracolsep{\fill}}lcccc}
\hline\hline
\multicolumn{1}{c}{Parameter}   &
\multicolumn{1}{c}{Close}       &
\multicolumn{1}{c}{Wide}        \\
\hline
$\chi^2$/dof            &  $1632.6/1585          $    &  $1594.3/1585          $     \\
$t_0$ (HJD$^\prime$)    &  $9405.996 \pm 0.051   $    &  $9405.747 \pm 0.093   $     \\
$u_0$ (10$^{-2}$)       &  $5.316 \pm 0.094      $    &  $5.156 \pm 0.278      $     \\
$\te$ (days)            &  $104.02 \pm 1.98      $    &  $105.53 \pm 5.38      $     \\
$s$                     &  $0.6866 \pm 0.0023    $    &  $1.5156 \pm 0.0230    $     \\
$q$                     &  $0.0790 \pm 0.0024    $    &  $0.0825 \pm 0.0055    $     \\
$\alpha$ (rad)          &  $0.7453 \pm 0.0058    $    &  $0.7186 \pm 0.0159    $     \\
$\rho$ ($10^{-3}$)      &  $0.284 \pm 0.044      $    &  $0.297 \pm 0.027      $     \\
\hline
\end{tabular*}
\end{table}

\subsection{KMT-2021-BLG-1643}\label{sec:four-three}

The lensing light curve of KMT-2021-BLG-1643 is presented in Figure~\ref{fig:three}. The event 
was found in its early stage by the KMTNet survey on 2021 June 8 ($\hjdp=9374$), at which the 
source was brighter than the baseline magnitude $I_{\rm base}=18.91$ by $\Delta I\sim 1.2$~mag. 
The source lies in the KMTNet BLG04 field, toward which the event was monitored with a 1~hr 
cadence. The event exhibited a pair of caustic spikes, which occurred at $\hjdp\sim ~9401.1$ and 
9403.4, and a weak bump, which was centered at $\hjdp\sim 9409$.  The region between the two 
caustic spikes exhibited a characteristic U-shape pattern, indicating that the spikes occurred 
when the source entered and exited a caustic. The first caustic spike was not resolved because 
the sky at the KMTA site was clouded out, but the second caustic was partially covered by the two 
KMTS and one KMTC data points.

From the 2L1S modeling of the light curve, we found a pair of solutions resulting from the 
close--wide degeneracy. The binary lensing parameters are $(s, q)_{\rm close}\sim (0.69, 0.08)$ 
and $(s, q)_{\rm wide}\sim (1.52, 0.08)$ for the close and wide solutions, respectively.  We 
list the full lensing parameters of the two solutions in Table~\ref{table:four}, and the model 
curves and residuals are presented in Figure~\ref{fig:three}.  From the comparison of the fits, 
it is found that the wide solution is preferred over the close solution by $\Delta\chi^2=38.3$, 
indicating that the degeneracy is lifted with a fairly strong confidence level.  Despite the fact 
that the caustic exit was partially covered by only a small number data points, the normalized 
source radius, $\rho\sim 0.3\times 10^{-3}$, could be constrained.

The measured event time scale, $\te\sim 105$~days, of the event comprises an important portion 
of a year, and thus it may be possible to constrain  microlens-parallax parameters.  We conducted 
an additional modeling considering the higher-order effects.  Figure~\ref{fig:four} shows the 
scatter  plot of points in the MCMC chain on the $\piee$--$\pien$ parameter plane.  It was found 
that the improvement of model fit with the inclusion of the higher-order effects is very minor, 
but the amplitude of the scatter plot provided a constraint on the physical lens parameters.

\begin{figure}[t]
\includegraphics[width=\columnwidth]{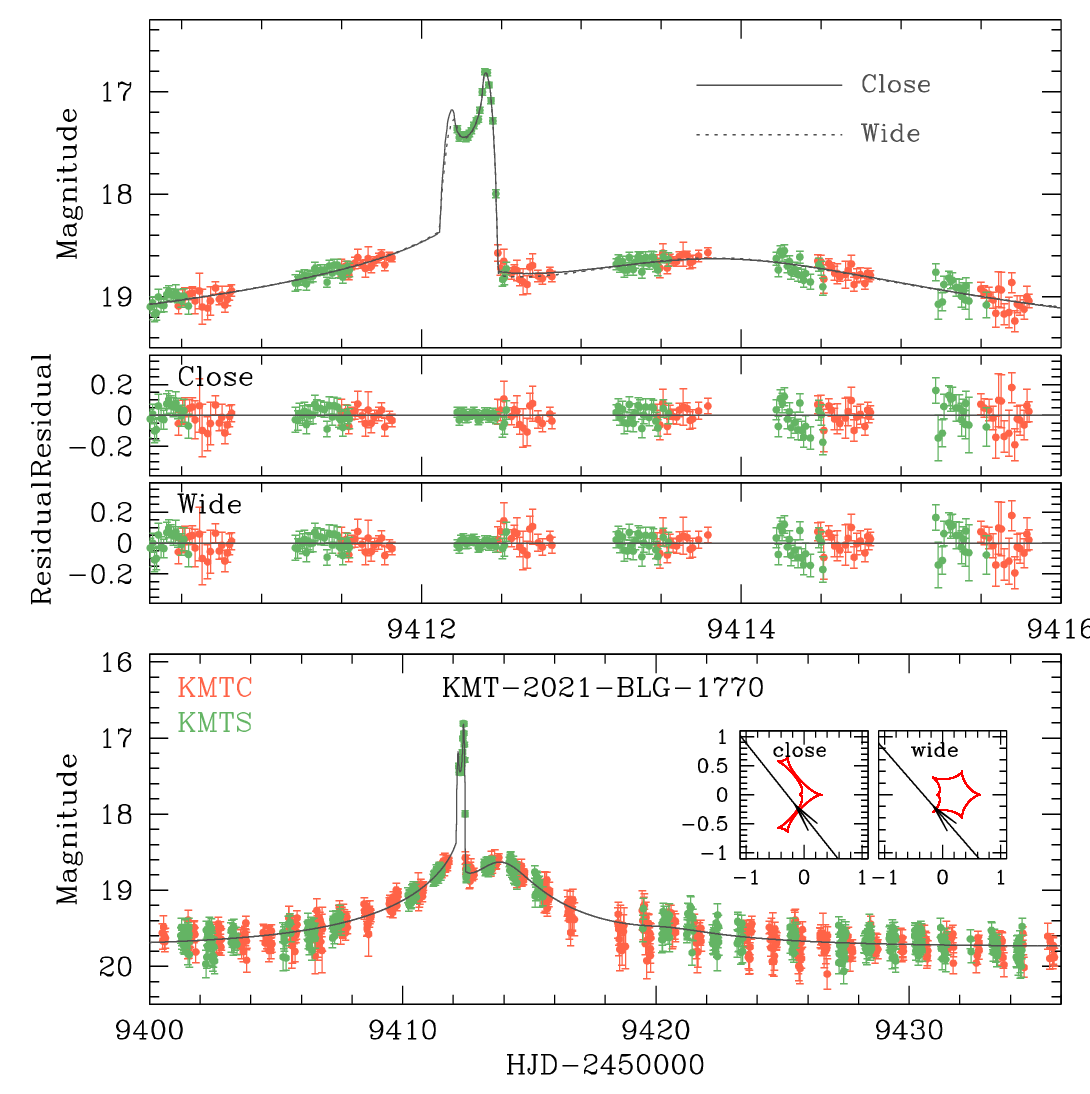}
\caption{
Light curve of KMT-2021-BLG-1770. 
The scheme and layout of the plot are same as those in Fig.~\ref{fig:one}.
}
\label{fig:five}
\end{figure}

We present the configurations of the close and wide lens systems in the two insets of the 
bottom panel of Figure~\ref{fig:three}. Similar to the case of KMT-2021-BLG-1110, the source 
passed the back-end side of the caustic. The spike features were produced by the source passage 
through the lower left cusp of the caustic, and the weak bump was generated by the source approach 
close to the left-side on-axis cusp of the caustic.

\begin{table}[t]
\small
\caption{Model parameters of KMT-2021-BLG-1770.\label{table:five}}
\begin{tabular*}{\columnwidth}{@{\extracolsep{\fill}}lcccc}
\hline\hline
\multicolumn{1}{c}{Parameter}   &
\multicolumn{1}{c}{Close}       &
\multicolumn{1}{c}{Wide}        \\
\hline
$\chi^2$/dof            &  $2895.9/2901         $     &  $ 2904.8/2901        $     \\
$t_0$ (HJD$^\prime$)    &  $9413.366 \pm 0.035  $     &  $ 9413.215 \pm 0.033 $     \\
$u_0$                   &  $0.227 \pm 0.014     $     &  $ 0.257 \pm 0.015    $     \\
$\te$ (days)            &  $7.57 \pm 0.31       $     &  $ 6.79 \pm 0.32      $     \\
$s$                     &  $0.8070 \pm 0.0045   $     &  $ 1.1403 \pm 0.0181  $     \\
$q$                     &  $0.1534 \pm 0.0063   $     &  $ 0.1889 \pm 0.0108  $     \\
$\alpha$ (rad)          &  $0.899 \pm 0.021     $     &  $ 0.860 \pm 0.021    $     \\
$\rho$ ($10^{-3}$)      &  $6.12 \pm 0.28       $     &  $ 7.03 \pm 0.34      $     \\
\hline
\end{tabular*}
\end{table}

\begin{table*}[t]
\footnotesize
\caption{Source properties and angular Einstein radii.\label{table:six}}
\begin{tabular}{lllll}
\hline\hline
\multicolumn{1}{c}{Quantity}            &
\multicolumn{1}{c}{KMT-2021-BLG-0588}   &
\multicolumn{1}{c}{KMT-2021-BLG-1110}   &
\multicolumn{1}{c}{KMT-2021-BLG-1643}   &
\multicolumn{1}{c}{KMT-2021-BLG-1770}   \\
\hline
$(V-I)_s$                   &  $1.259 \pm 0.021  $  &  $2.674 \pm 0.037 $   &   $2.129 \pm 0.039 $  &  $2.485 \pm 0.083 $  \\
$I_s$                       &  $19.382 \pm 0.010 $  &  $21.866 \pm 0.026$   &   $22.215 \pm 0.011$  &  $20.005 \pm 0.018$  \\
$(V-I, I)_b$                &  $(0.927, 18.708)  $  &  $(2.684, 20.759) $   &   $(2.518, 19.536) $  &  $(2.585, 18.924) $  \\
$(V-I, I)_{\rm RGC}$        &  $(1.660, 15.873)  $  &  $(2.914, 16.869) $   &   $(2.212, 16.034) $  &  $(2.723, 16.522) $  \\
$(V-I, I)_{{\rm RGC},0}$    &  $(1.060, 14.336)  $  &  $(1.060, 14.452) $   &   $(1.060, 14.423) $  &  $(1.060, 14.348) $  \\
$(V-I)_{s,0}$               &  $0.660 \pm 0.021, $  &  $0.820 \pm 0.037 $   &   $0.997 \pm 0.039 $  &  $0.822 \pm 0.083 $  \\
$I_{s,0}$                   &  $18.294 \pm 0.010 $  &  $19.450 \pm 0.026$   &   $20.604 \pm 0.011$  &  $17.830 \pm 0.018$  \\
 Source type                &   G0V                 &    G9V                &    K3V                &   G9V                \\
$\theta_*$ ($\mu$as)        &  $0.651 \pm 0.048  $  &  $0.458 \pm 0.036 $   &   $0.323 \pm 0.026 $  & $0.969 \pm 0.105  $  \\
$\thetae$ (mas)             &  $0.904 \pm 0.084  $  &  $0.578 \pm 0.051 $   &   $1.085 \pm 0.170 $  & $0.158 \pm 0.019  $  \\
\hline
\end{tabular}
\end{table*}

\begin{figure}[t]
\includegraphics[width=\columnwidth]{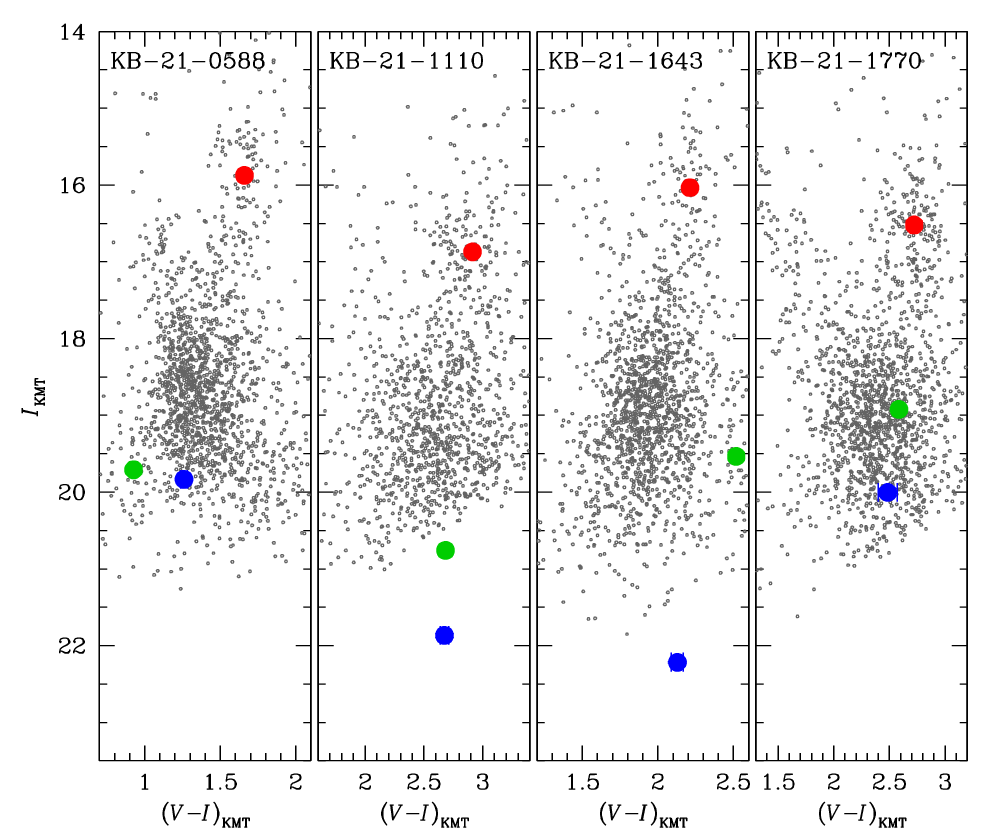}
\caption{
Source positions (blue dots) in the instrumental color-magnitude diagrams of stars lying near
the source stars of the individual events. The red and green dots in each panel represent the 
centroid of the red giant clump and position of the blend, respectively.
}
\label{fig:six}
\end{figure}

\subsection{KMT-2021-BLG-1770}\label{sec:four-four}

Figure~\ref{fig:five} shows the light curve of the lensing event KMT-2021-BLG-1770. The event 
was found by the KMTNet group on 2021 July 16 ($\hjdp\sim 9406$). The source, which had a 
baseline magnitude $I_{\rm base}=19.06$, was in the KMTNet prime field BLG03, for which images 
were taken with a 0.5~hr cadence.  Most region of this field overlaps with the region covered by 
the BLG43 field, but the event lies in the offset region that was not covered by the BLG43 field. 
In our analysis, we do not use the KMTA data set due to its low photometric quality. Similar to 
the event KMT-2021-BLG-1643, the light curve of KMT-2021-BLG-1770 is characterized by a pair of 
caustic spikes and a following weak bump.  The first caustic spike, which occurred at 
$\hjdp=9412.2$, was not covered, but the second spike, which occurred at $\hjdp=9412.4$, and 
the U-shape region between the two spikes were resolved by the combination of the KMTS and KMTC 
data sets.  The weak bump is centered at $\hjdp\sim 9414$, which was about 2~days after the caustic 
spikes.

From the analyses of the light curve, we identified two local solutions, in which one solution 
has a binary separation $s<1$ (close solution) and the other has a separation $s>1$ (wide solution).  
The model curves of the solutions are drawn over the data points and residuals from the models are 
shown in Figure~\ref{fig:five}. The binary lensing parameters of the individual solutions are 
$(s, q)_{\rm close}\sim (0.81, 0.15)$ and $(s, q)_{\rm wide}\sim (1.14, 0.19)$. As stated, the 
event was chosen as a BD candidate despite the fact that the mass ratio between the lens components 
is slightly greater than the threshold mass ratio $q_{\rm th}=0.1$, because the event time scale, 
$\te\sim 7$~days, is substantially shorter than several-week time scale of typical lensing events.  
The normalized source radius, $\rho\sim (6-7)\times 10^{-3}$, was measured from analyzing the 
caustic-exit part of the light curve.

The lens-system configurations of the close and wide solutions are presented in the two insets 
of the bottom panel of Figure~\ref{fig:five}. It is found that the configurations of the close 
and wide solutions are very similar to those of the corresponding solutions of KMT-2021-BLG-0588. 
That is, the caustic spikes were generated by the passage of the source through the slim bridge 
part connecting the central and peripheral caustics according to the close solution, and by the 
source pass through the tip of the lower left cusp of the six-sided resonant caustic according 
to the wide solution. The difference between the solutions of the two events is that the close 
solution is preferred over the wide solution by $\Delta\chi^2=8.9$ in the case of KMT-2021-BLG-1770, 
while the wide solution yields a better fit than the close solution in the case of KMT-2021-BLG-0588. 
For the same reason mentioned in Sect.~\ref{sec:four-one}, the similarity between the model curves 
of the close and wide solutions is caused by an accidental degeneracy rather than a close--wide 
degeneracy.

\section{Source stars and Einstein radii}\label{sec:five}

In this section, we specify the source stars of the events. Specifying the source star of a
caustic-crossing 2L1S event is important to estimate the angular Einstein radius from the relation
\begin{equation}
\thetae = {\theta_*\over \rho},
\label{eq3}
\end{equation}
where the normalized source radius $\rho$ is measured by analyzing the caustic-crossing parts of 
the light curve, and the angular source radius $\theta_*$ can be deduced from the source type.

\begin{table}[t]
\small
\caption{Astrometric centroid offsets.\label{table:seven}}
\begin{tabular*}{\columnwidth}{@{\extracolsep{\fill}}lcccc}
\hline\hline
\multicolumn{1}{c}{Event}                       &
\multicolumn{1}{c}{$\delta\theta$ (mas)}        \\
\hline
KMT-2021-BLG-0588  &  $248.83 \pm 10.22$   \\
KMT-2021-BLG-1110  &  $1110.87 \pm 8.03$   \\
KMT-2021-BLG-1643  &  $172.38 \pm 8.01 $   \\
KMT-2021-BLG-1770  &  $373.34 \pm 10.19$   \\
\hline
\end{tabular*}
\end{table}

We specified the source stars of the individual events by measuring their de-reddened colors and 
magnitudes. To estimate the de-reddended color and magnitude, $(V-I, I)_0$, from the instrumental 
values, $(V-I, I)_s$, we applied the \citet{Yoo2004} method, in which the centroid of red giant 
clump (RGC) is used as a reference for the calibration. Following the routine procedure of the 
method, we first estimated instrumental $I$ and $V$-band magnitudes of the source by regressing 
the photometry data of the individual passbands processed using the pyDIA code, and placed the 
source in the instrumental CMD of stars around the source constructed using the same pyDIA code. 
We then measured the offsets in color and magnitude, $\Delta (V-I, I)$, of the source from the 
RGC centroid, and estimated de-reddened color and magnitude as 
\begin{equation}
(V - I, I)_{s,0} = (V - I, I)_{{\rm RGC},0} + \Delta (V - I, I),
\label{eq4}
\end{equation}
where $(V - I, I)_{{\rm RGC},0}$ are the de-reddened color and magnitude of the RGC centroid 
known from \citet{Bensby2013} and \citet{Nataf2013}, respectively.

Figure~\ref{fig:six} shows the positions of the source (blue dot) and RGC centroid (red dot) 
in the instrumental CMDs of the individual events. In Table~\ref{table:six}, we list the 
values of $(V-I, I)_s$, $(V-I, I)_{\rm RGC}$, $(V-I, I)_{{\rm RGC},0}$. and $(V-I, I)_{s,0}$ 
estimated from the procedure described in the previous paragraph. According to the estimated 
colors and magnitudes, the spectral types of the source stars are G0V, G9V, K3V, and G9V for 
KMT-2021-BLG-0588, KMT-2021-BLG-1110, KMT-2021-BLG-1643, and KMT-2021-BLG-1770, respectively. 
With the measured source color and magnitude, we estimated the angular radius of source star 
by first converting $V-I$ color into $V-K$ color using the \citet{Bessell1988} relation, and 
then by deducing $\theta_*$ from the \citet{Kervella2004} relation between $(V-K, V)$ and 
$\theta_*$. With the measured source radii, the angular Einstein radii were estimated using 
the relation in Equation~(\ref{eq3}).  We list the the estimated values of  $\theta_*$ and 
$\thetae$ of the individual events in the bottom two lines of Table~\ref{table:six}.

\begin{table*}[t]
\footnotesize
\caption{Physical lens parameters.\label{table:eight}}
\begin{tabular}{lllll}
\hline\hline
\multicolumn{1}{c}{Parameter}           &
\multicolumn{1}{c}{KMT-2021-BLG-0588}   &
\multicolumn{1}{c}{KMT-2021-BLG-1110}   &
\multicolumn{1}{c}{KMT-2021-BLG-1643}   &
\multicolumn{1}{c}{KMT-2021-BLG-1770}   \\
\hline 
$M_1$ ($M_\odot$)     &  $0.54^{+0.31}_{-0.24}    $  &  $0.74^{+0.27}_{-0.35}   $   &   $0.73^{+0.24}_{-0.17}   $  &  $0.13^{+0.18}_{-0.07}   $  \\ [0.8ex]
$M_2$ ($M_\odot$)     &  $0.053^{+0.031}_{-0.023} $  &  $0.055^{+0.020}_{-0.026}$   &   $0.061^{+0.020}_{-0.014}$  &  $0.020^{+0.028}_{-0.011}$  \\ [0.8ex]
$\dl$ (kpc)           &  $3.03^{+0.94}_{-1.00}    $  &  $5.91^{+0.94}_{-1.52}   $   &   $3.35^{+0.97}_{-0.77}   $  &  $6.92^{+0.97}_{-1.01}   $  \\ [0.8ex]
$a_\perp$ (AU)        &  $3.38^{+0.96}_{-1.02}    $  &  $8.46^{+1.34}_{-2.15}   $   &   $5.36^{+1.46}_{-1.17}   $  &  $0.90^{+0.13}_{-0.13}   $  \\ [0.8ex]
\hline
\end{tabular}
\end{table*}

\begin{figure}[t]
\includegraphics[width=\columnwidth]{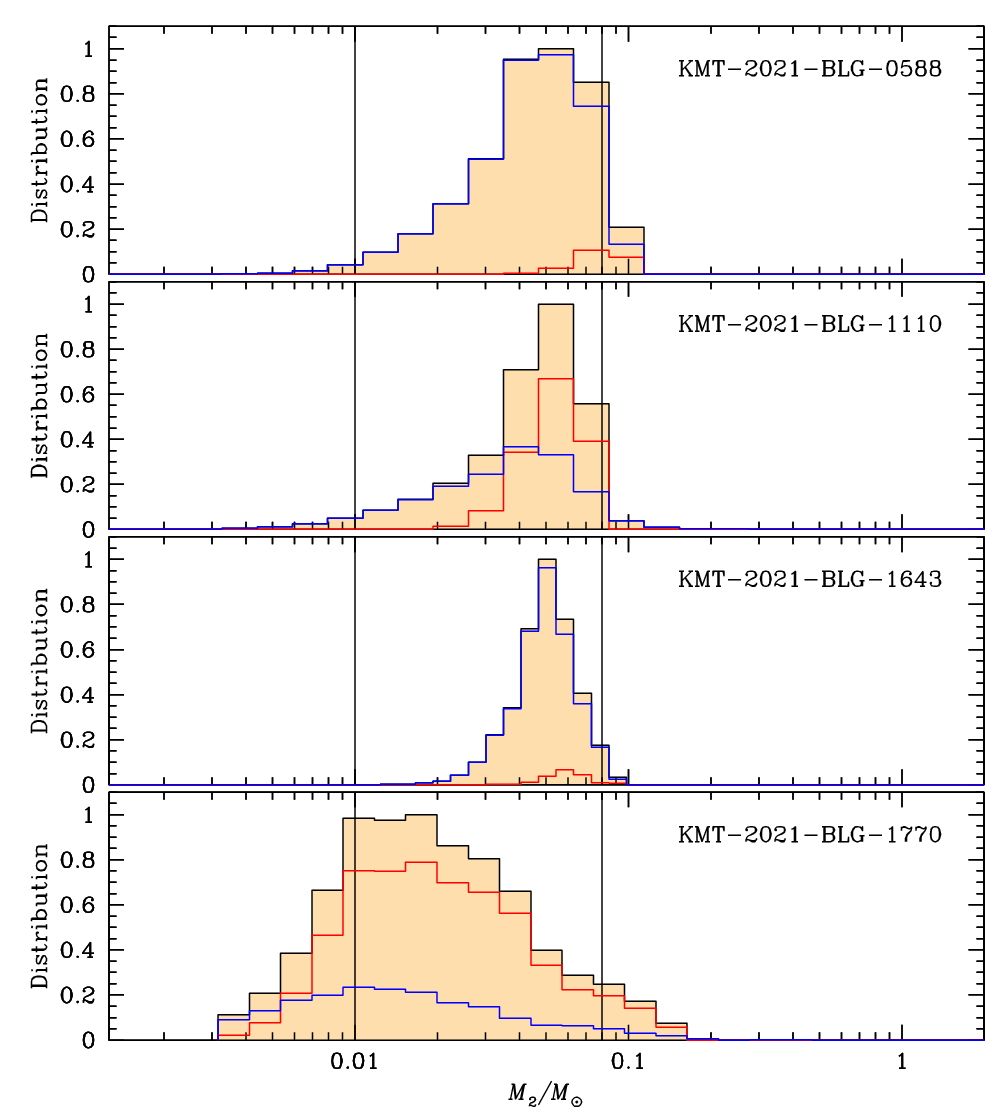}
\caption{
Bayesian posteriors for the companion mass of the lens. In each panel, the two vertical solid line
represent the mass range of brown dwarf. The blue and red curves indicate the contributions by the 
disk and bulge lens populations, respectively. 
}
\label{fig:seven}
\end{figure}

Also marked in Figure~\ref{fig:six} are the positions of the blend (green dots) in the CMDs 
of the individual events. We list the measured values of the color and magnitude of the blend, 
$(V-I, I)_b$, in Table~\ref{table:six}. Besides KMT-2021-BLG-0588, for which the blended light 
is similar to the flux of the source, it is found that the blended fluxes are substantially 
greater than the source fluxes. In order to check the possibility that the lens is the main 
origin of the blended flux, we measured the astrometric offset $\delta\theta$ between the 
centroid of the source measured at the peak time of the lensing magnification and that measured 
at the baseline.  If the lens were the main origin of the blended flux, the offset would be very 
small because the relative lens-source proper motions are $< 10$~mas/yr for all events. In the 
case that the origin of the blended flux is a nearby star, which is typically separated from 
the source by an order of 100~mas, the resulting astrometric offset would be substantially 
greater than the typical astrometric precision of order 10~mas. In Table~\ref{table:seven}, we 
list the measured centroid offsets of the individual events.  For all events, it is found that 
the astrometric offsets are much greater than the measurement precision, and this indicates 
that the origins of the blended light are nearby stars rather than the lenses.

\begin{figure}[t]
\includegraphics[width=\columnwidth]{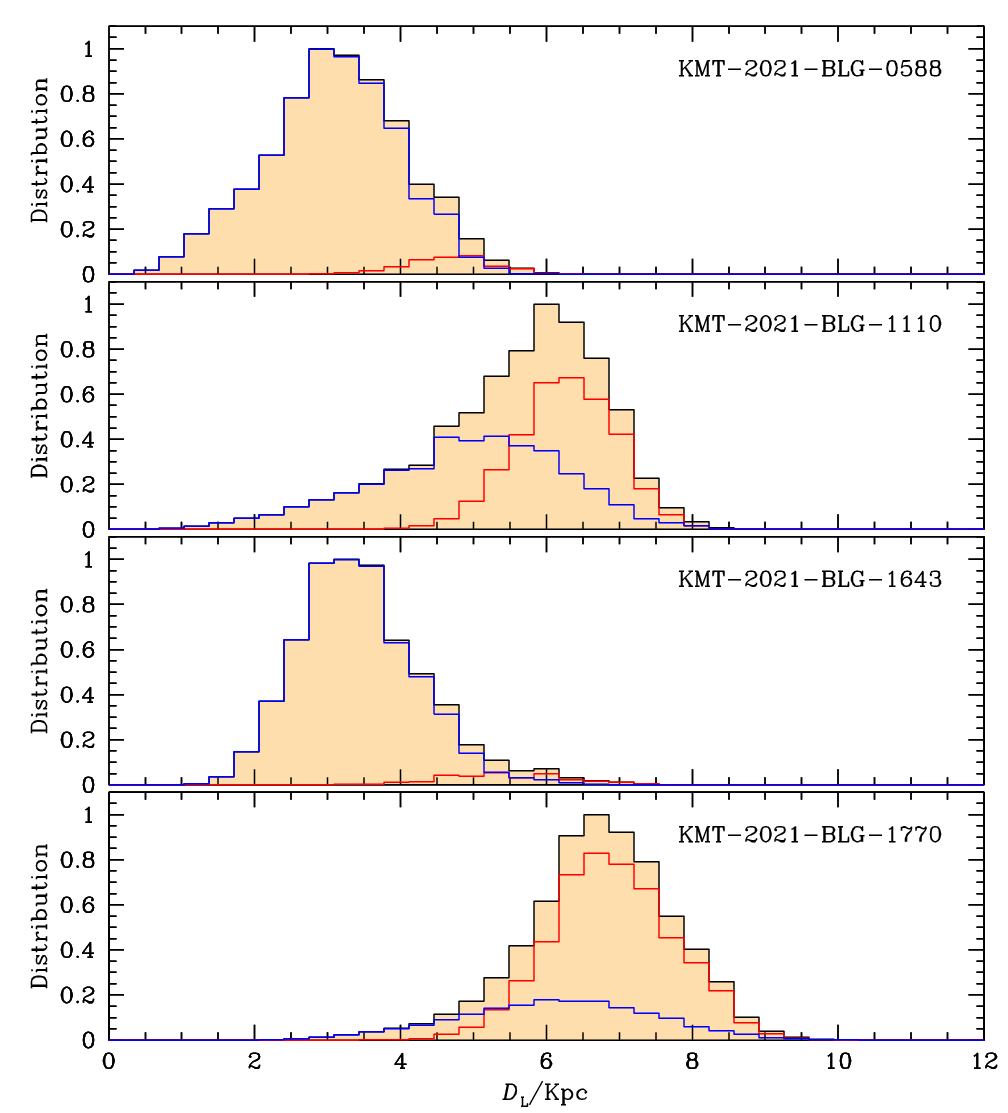}
\caption{
Bayesian posteriors for the distance to the lens. The layout of the plots are same as those in
Fig.~\ref{fig:seven}. 
}
\label{fig:eight}
\end{figure}

\section{Physical lens parameters}\label{sec:six}

The mass $M$ and distance $\dl$ to the lens can be constrained by measuring lensing observables:
$\te$, $\thetae$, and $\pie$. The event time scale is the basic observable that is measurable 
for general lensing events, and the angular Einstein radius is another observable that is measurable 
for events with light curves affected by finite-source effects. These two observables are related 
to the physical lens parameters by the relations in Equation~(\ref{eq1}). With the measurement of 
the extra observable $\pie$, the physical lens parameters would be uniquely determined from the 
relations in Equation~(\ref{eq2}). For the analyzed events, the observables $\te$ and $\thetae$ were 
measured, but $\pie$ was not securely measured for any of the events. Without the constraint of $\pie$, 
we estimated the physical lens parameters by conducting Bayesian analyses of the events using models 
of physical and dynamical distributions and mass function of objects in our Galaxy together with the 
constraints provided by the measured blended flux.

In the first step of the Bayesian analysis, we conducted a Monte Carlo simulation to generate 
a large number of artificial lensing events. For each artificial event, the distances of the 
lens and source and their relative proper motion were assigned using a Galactic model, and the 
mass of the lens was assigned using a model mass function. In the simulation, we adopted the 
Galactic model of \citet{Jung2021} and the mass function model of \citet{Jung2018}.  In the mass 
function, we included white-dwarf remnants but exclude black holes and neuron stars.  In the 
second step, we computed the lensing observables $(t_{{\rm E},i}, \theta_{{\rm E},i})$ corresponding 
to the assigned values $(M, \dl, \ds, \mu)$ of each artificial event using the relations in 
Equation~(\ref{eq1}). In the final step, we constructed Bayesian posteriors of the lens mass and 
distance by imposing a weight $w_i=\exp(-\chi^2/2)$ on each event.  Here the $\chi^2$ value was 
calculated as 
\begin{equation}
\chi_i^2 = 
\left[ {t_{{\rm E},i}-\te \over \sigma(\te)}\right]^2 +
\left[ {\theta_{{\rm E},i}-\thetae \over \sigma(\thetae)}\right]^2,
\label{eq5}
\end{equation}
where $[\te, \sigma(\te)]$ and $[\thetae, \sigma(\thetae)]$ represent the measured values and 
uncertainties of the observables $\te$ and $\thetae$, respectively.  For the event KMT-2021-BLG-1643 
with a long event time scale, we imposed the $\pie$ constraint by including an additional term 
$\sum_{j=1}^2 \sum_{k=1}^2 b_{j,k} (\pi_{{\rm E},j,i}-\pi_{{\rm E},i}) (\pi_{{\rm E},k,i}-
\pi_{{\rm E},i})$ to the right side of Eq.~(\ref{eq5}).

Besides the constraints from the lensing observables, we additionally imposed the blending
constraint in the Bayesian analyses. This constraint is provided by the fact that the flux 
from the lens comprises a portion of the total blending flux, and thus the lens flux should 
be less than the total blending flux. For the imposition of this constraint, we calculated 
the lens brightness as 
\begin{equation}
I_L = M_{I,L} + 5 \log \left({\dl\over {\rm pc}}\right) - 5 + A_{I,{\rm tot}},
\label{eq6}
\end{equation}
where $M_{I,L}$ denotes the absolute $I$-band magnitude corresponding to the lens mass, and 
$A_{I,L}$ is the extinction to the lens lying at a distance $\dl$. The extinction was modeled as 
\begin{equation}
A_{I,L} = A_{I,{\rm tot}} \left[ 1-\exp\left( -{|z|\over h_{z,{\rm dust}}}\right)\right],
\label{eq7}
\end{equation}
where $A_{I,{\rm tot}}$ denotes the total extinction toward the field, $h_{z,{\rm dust}} = 
100$~pc is the adopted vertical scale height of dust, $z = \dl \sin b + z_0$ and $z_0=15$~pc 
represent the vertical positions of the lens and the sun above the Galactic plane, respectively. 
The values $A_{I,{\rm tot}}$ for the individual events are listed in Table~\ref{table:one}. It 
was found that the blending constraint had important effects on the determined physical parameters 
of the events KMT-2012-BLG-0558 and KMT-2021-BLG-1643, for which the lenses are expected to be 
located relatively nearby to the Sun based on their large Einstein  radii.  Below we discuss 
this issue in more detail.

In Figures~\ref{fig:seven} and \ref{fig:eight}, we present the Bayesian posteriors of the mass of 
the binary lens companion and distance to the lens system, respectively. The estimated values of 
the primary ($M_1$) and companion ($M_2$) masses, distance, and projected separation between the 
lens components ($a_\perp=s\thetae\dl$) are listed in Table~\ref{table:eight}.  For each parameter, 
the median value was adopted as a representative value and the upper and lower ranges of the 
uncertainty were chosen as the 16\% and 84\% of the posterior distribution, respectively.  
According to the estimated masses, it is found that the masses of the lens companions are well 
within the BD mass range $0.012<M_2/M_\odot\leq 0.076$ (or $13<M_2/M_{\rm J}\leq 80$), although 
there is some variation of the primary masses, which lie in the mass range of main-sequence stars 
with spectral types from K to M.  In Table~\ref{table:nine}, we list the probabilities for the 
companions of the individual lenses being in the BD mass range, $P_{\rm BD}$. It is found that 
the probabilities are greater than 59\% in all cases of the events.  
For KMT-2021-BLG-1770L, the mass of the primary is so small that it can be a BD as 
well with a probability of $P_{\rm BD}\sim 35\%$. 
In this case, the lens is a BD binary like 
OGLE-2009-BLG-151L, OGLE-2011-BLG-0420L \citep{Choi2013}
OGLE-2016-BLG-1266L \citep{Albrow2018},
OGLE-2016-BLG-1469L \citep{Han2017}, 
MOA-2016-BLG-231L \citep{Chung2019}, and
OGLE-2017-BLG-1038L \citep{Malpas2022}.

\begin{table}[t]
\small
\caption{BD, disk, and bulge lens probabilities.\label{table:nine}}
\begin{tabular*}{\columnwidth}{@{\extracolsep{\fill}}lcccc}
\hline\hline
\multicolumn{1}{c}{Event}                     &
\multicolumn{1}{c}{$P_{\rm BD}$ (\%)}         &
\multicolumn{1}{c}{$P_{\rm disk}$ (\%)}       &
\multicolumn{1}{c}{$P_{\rm bulge}$ (\%)}      \\
\hline
 KMT-2021-BLG-0588     &   82     &   95    &   5   \\
 KMT-2021-BLG-1110     &   85     &   53    &   47  \\
 KMT-2021-BLG-1643     &   91     &   95    &   5   \\
 KMT-2021-BLG-1770     &   59     &   24    &   76  \\
\hline
\end{tabular*}
\end{table}

\begin{table*}[t]
\footnotesize
\caption{Relative proper motion, angular separation in 2030, and $K$-band source magnitude.\label{table:ten}}
\begin{tabular}{lllll}
\hline\hline
\multicolumn{2}{c}{Event}                         &
\multicolumn{1}{c}{$\mu$ (mas/yr)}                &
\multicolumn{1}{c}{$\Delta\theta_{2030}$ (mas)}   &
\multicolumn{1}{c}{K (mag)}                       \\
\hline
Paper I    &  OGLE-2016-BLG-0890  &  $6.30 \pm 1.12$   &  $88.2 \pm 15.68  $   &  $12.44 \pm 0.16 $  \\
           &  MOA-2017-BLG-477    &  $9.33 \pm 0.83$   &  $121.29 \pm 10.79$   &  $18.21 \pm 0.10 $  \\
           &  OGLE-2017-BLG-0614  &    --              &     --                &  $19.08 \pm 0.12 $  \\
           &  KMT-2018-BLG-0357   &  $7.52 \pm 1.05$   &  $90.24 \pm 12.6  $   &  $17.81 \pm 0.12 $  \\
           &  OGLE-2018-BLG-1489  &  $4.89 \pm 0.36$   &  $58.68 \pm 4.32  $   &  $16.50 \pm 0.02 $  \\
           &  OGLE-2018-BLG-0360  &  $4.12 \pm 0.59$   &  $49.44 \pm 7.08  $   &  $16.83 \pm 0.13 $  \\
\hline
Paper II   &  KMT-2018-BLG-0321   &  $> 2.4        $   &  $> 28.8         $    &  $16.05 \pm 0.11 $  \\
           &  KMT-2018-BLG-0885   &  $> 1.6        $   &  $> 19.2         $    &  $18.58 \pm 0.13 $  \\
           &  KMT-2019-BLG-0297   &  $6.93 \pm 0.51$   &  $76.23 \pm 5.61 $    &  $17.88 \pm 0.01 $  \\
           &  KMT-2019-BLG-0335   &  $2.87 \pm 1.16$   &  $31.57 \pm 12.76$    &  $18.89 \pm 0.17 $  \\
\hline                                                                        
This work  &  KMT-2021-BLG-0588   &  $8.68 \pm 0.80$   &  $78.12 \pm 7.20$     &  $17.68 \pm 0.02 $  \\
           &  KMT-2021-BLG-1110   &  $7.08 \pm 0.62$   &  $63.72 \pm 5.58$     &  $18.75 \pm 0.05 $  \\
           &  KMT-2021-BLG-1643   &  $3.76 \pm 0.58$   &  $33.84 \pm 5.22$     &  $19.51 \pm 0.04 $  \\
           &  KMT-2021-BLG-1770   &  $7.63 \pm 0.90$   &  $68.67 \pm 8.10$     &  $17.08 \pm 0.09 $  \\
\hline
\end{tabular}
\end{table*}

In Table~\ref{table:nine}, we list the probabilities of the lenses being in the disk, $P_{\rm disk}$, 
and bulge, $P_{\rm bulge}$. For the events KMT-2021-BLG-0588 and KMT-2021-BLG-1643, it is very likely 
that the lenses lie in the disk, while the lens of KMT-2021-BLG-1770 is likely to lie in the bulge. 
For KMT-2021-BLG-1110, on the other hand, the disk and bulge probabilities are approximately the 
same. It is found that the constraint on the lens location comes mainly from the estimated radius 
of the Einstein ring. For the events KMT-2021-BLG-0588 and KMT-2021-BLG-1643, the respective 
Einstein radii are $\thetae\sim 0.90$~mas and $\sim 1.08$~mas, which are approximately two times 
bigger than the typical Einstein radius of $\sim 0.5$~mas for the event produced by a low-mass 
stellar lens with a mass $M\sim 0.3~M_\odot$ lying about halfway between the sun and a bulge 
source. By contrast, the Einstein radius $\thetae\sim 0.16$~mas of KMT-2022-BLG-1770 is 
substantially smaller than the typical value, and thus $P_{\rm bulge}$ is substantially higher 
than $P_{\rm disk}$. The Einstein radius $\thetae\sim 0.58$~mas of KMT-2021-BLG-1110 is close 
to the typical value, and thus $P_{\rm disk}$ and $P_{\rm bulge}$ are approximately the same. 
In the posterior distributions presented in Figures~\ref{fig:seven} and \ref{fig:eight}, we mark 
the contributions of the disk and bulge lens populations by blue and red curves, respectively.

\section{Summary and discussion }\label{sec:seven}

Following the works in papers~I and II, we reported the BD companions in binary lenses found
from the inspection of the microlensing data collected in the 2021 season by the high-cadence
surveys, including KMT-2021-BLG-0588LB, KMT-2021-BLG-1110LB, KMT-2021-BLG-1643LB, and
KMT-2021-BLG-1770LB. Modeling the light curve of each event yielded a pair of solutions with
projected separations smaller and greater than the Einstein radius, but the degeneracy between 
the solutions was resolved with a strong confidence level except for KMT-2021-BLG-1770, for which 
the resolution of the degeneracy was less clear than the others.  From the Bayesian analyses 
conducted with the constraints provided by the observables of the event time scale and Einstein 
radius together with the constraint from the blended light, it was estimated that the masses of 
the primary and companion of the individual events are 
$(M_1/M_\odot, M_2/M_\odot)= 
(0.54^{+0.31}_{-0.24}, 0.053^{+0.031}_{-0.023})$ for KMT-2021-BLG-0588L, 
$(0.74^{+0.27}_{-0.35}, 0.055^{+0.020}_{-0.026})$ for KMT-2021-BLG-1110L, 
$(0.73^{+0.24}_{-0.17}, 0.061^{+0.020}_{-0.014})$ for KMT-2021-BLG-1643L, and 
$(0.13^{+0.18}_{-0.07}, 0.020^{+0.028}_{-0.011})$ for KMT-2021-BLG-1770L.  
The estimated masses of the binary companions were well within the BD mass range, although there 
was some variation of the primary masses, which were in the mass range of main-sequence stars 
with spectral types from K to M. The probabilities of the lens companions being in the BD mass 
range were estimated as 82\%, 85\%, 91\%, and 59\% for the individual events.

The BD nature of the lens companions presented in this work and papers~I and II can be 
confirmed by directly imaging the lenses from future high-resolution adaptive-optics (AO) 
followup observations when the lenses are separated from the source stars \citep{Gould2022a}. 
For these followup observations, we compute the lens-source separations $\Delta\theta_{2030}$ 
expected in 2030, which is an approximate year of the first AO light on 30~m class telescopes. 
In Table~\ref{table:ten}, we list the relative lens-source proper motions, expected lens-source 
separations, and $K$-band source magnitudes of the BD events reported in this work and papers~I 
and II. The $K$-band source magnitude was estimated as $K = I_{s,0} + (V-I)_0 - (V-K)_0 + A_I/7$, 
and the separation is estimated as $\Delta\theta_{2030} =\mu\Delta t$, where the relative 
lens-source proper motion is computed by $\mu=\thetae/\te$ and $\Delta t$ indicates the time 
gap between the peak of the event and the year 2030. We note that $\Delta\theta_{2030}$ of the 
event OGLE-2017-BLG-0614 is not listed because the Einstein radius and the resulting proper 
motion could not be measured, and only the lower limits are listed for KMT-2018-BLG-0321 and 
KMT-2018-BLG-0885 because only the lower limits of $\thetae$ were constrained for these events.  
From the table, one finds that the separations are greater than 30~mas for all events with 
measured proper motions, and except for the two events KMT-2019-BLG-0335 and KMT-2021-BLG-1643, 
the separations are greater than $\sim 50$~mas, which will be adequate for the clear resolution 
of the lens from the source.  By comparing the relative lens-source proper motion estimated from 
the model with the value measured from followup AO observations, one can confirm the solution.  
Furthermore, from the stellar type of the primary lens, which comprises most of the flux from the 
lens, the approximate mass of the lens can be estimated.  This together with the estimated mass 
ratio enables one to confirm the BD nature of the lens companion.  We note that this test of 
presented solutions will be most useful for events with relative accuracy of relative proper 
motion better than 10\%.

\begin{acknowledgements}
Work by C.H. was supported by the grants of National Research Foundation of Korea 
(2019R1A2C2085965).
This research has made use of the KMTNet system operated by the Korea Astronomy and Space
Science Institute (KASI) at three host sites of CTIO in Chile, SAAO in South Africa, and 
SSO in Australia. Data transfer from the host site to KASI was supported by the Korea Research 
Environment Open NETwork (KREONET). 
This research was supported by the Korea Astronomy and Space Science Institute under the R\&D
program (Project No. 2023-1-832-03) supervised by the Ministry of Science and ICT.
The MOA project is supported by JSPS KAKENHI Grant Number JP24253004, JP26247023, JP23340064, 
JP15H00781, JP16H06287, JP17H02871 and JP22H00153.
J.C.Y., I.G.S., and S.J.C. acknowledge support from NSF Grant No. AST-2108414. 
Y.S.  acknowledges support from BSF Grant No 2020740.
\end{acknowledgements}

\end{document}